\documentstyle[12pt]{article}

\catcode`\@=11
\long\def\@makefntext#1{
\protect\noindent \hbox to 3.2pt {\hskip-.9pt
$^{{\ninerm\@thefnmark}}$\hfil}#1\hfill}                

 \def\@makefnmark{\hbox to 0pt{$^{\@thefnmark}$\hss}}  

\def\ps@myheadings{\let\@mkboth\@gobbletwo
\def\@oddhead{\hbox{}
\rightmark\hfil\ninerm\thepage}
\def\@oddfoot{}\def\@evenhead{\ninerm\thepage\hfil
\leftmark\hbox{}}\def\@evenfoot{}
\def\sectionmark##1{}\def\subsectionmark##1{}}

\newcounter{sectionc}\newcounter{subsectionc}\newcounter{subsubsectionc}
\renewcommand{\section}[1] {\vspace{0.6cm}\addtocounter{sectionc}{1}
\setcounter{subsectionc}{0}\setcounter{subsubsectionc}{0}\noindent
        {\bf\thesectionc. #1}\par\vspace{0.4cm}}
\renewcommand{\subsection}[1] {\vspace{0.6cm}\addtocounter{subsectionc}{1}
        \setcounter{subsubsectionc}{0}\noindent
        {\it\thesectionc.\thesubsectionc. #1}\par\vspace{0.4cm}}
\renewcommand{\subsubsection}[1]
{\vspace{0.6cm}\addtocounter{subsubsectionc}{1}
        \noindent {\rm\thesectionc.\thesubsectionc.\thesubsubsectionc.
        #1}\par\vspace{0.4cm}}
\newcommand{\nonumsection}[1] {\vspace{0.6cm}\noindent{\bf #1}
        \par\vspace{0.4cm}}

\newcounter{appendixc}
\newcounter{subappendixc}[appendixc]
\newcounter{subsubappendixc}[subappendixc]

\renewcommand{\appendix}[1] {\vspace{0.6cm}
        \refstepcounter{appendixc}
        \setcounter{figure}{0}
        \setcounter{table}{0}
        \setcounter{equation}{0}
        \renewcommand{\thefigure}{\Alph{appendixc}.\arabic{figure}}
        \renewcommand{\thetable}{\Alph{appendixc}.\arabic{table}}
        \renewcommand{\theappendixc}{\Alph{appendixc}}
        \renewcommand{\theequation}{\Alph{appendixc}.\arabic{equation}}
        \noindent{\bf Appendix \theappendixc #1}\par\vspace{0.4cm}}

\def\abstracts#1{{
        \centering{\begin{minipage}{30pc}\tenrm\baselineskip=12pt\noindent
        \centerline{\tenrm ABSTRACT}\vspace{0.3cm}
        \parindent=0pt #1
        \end{minipage}}\par}}

\renewenvironment{thebibliography}[1]
        {\begin{list}{\arabic{enumi}.}
        {\usecounter{enumi}\setlength{\parsep}{0pt}
\setlength{\leftmargin 1.25cm}{\rightmargin 0pt}
         \setlength{\itemsep}{0pt} \settowidth
        {\labelwidth}{#1.}\sloppy}}{\end{list}}

\newcounter{itemlistc}
\newcounter{romanlistc}
\newcounter{alphlistc}
\newcounter{arabiclistc}

\newcommand{\fcaption}[1]{
        \refstepcounter{figure}
        \setbox\@tempboxa = \hbox{\tenrm Fig.~\thefigure. #1}
        \ifdim \wd\@tempboxa > 6in
           {\begin{center}
        \parbox{6in}{\tenrm\baselineskip=12pt Fig.~\thefigure. #1}
            \end{center}}
        \else
             {\begin{center}
             {\tenrm Fig.~\thefigure. #1}
              \end{center}}
        \fi}

\newcommand{\tcaption}[1]{
        \refstepcounter{table}
        \setbox\@tempboxa = \hbox{\tenrm Table~\thetable. #1}
        \ifdim \wd\@tempboxa > 6in
           {\begin{center}
        \parbox{6in}{\tenrm\baselineskip=12pt Table~\thetable. #1}
            \end{center}}
        \else
             {\begin{center}
             {\tenrm Table~\thetable. #1}
              \end{center}}
        \fi}

\def\@citex[#1]#2{\if@filesw\immediate\write\@auxout
        {\string\citation{#2}}\fi
\def\@citea{}\@cite{\@for\@citeb:=#2\do
        {\@citea\def\@citea{,}\@ifundefined
        {b@\@citeb}{{\bf ?}\@warning
        {Citation `\@citeb' on page \thepage \space undefined}}
        {\csname b@\@citeb\endcsname}}}{#1}}

\newif\if@cghi
\def\cite{\@cghitrue\@ifnextchar [{\@tempswatrue
        \@citex}{\@tempswafalse\@citex[]}}
\def\citelow{\@cghifalse\@ifnextchar [{\@tempswatrue
        \@citex}{\@tempswafalse\@citex[]}}
\def\@cite#1#2{{$\null^{#1}$\if@tempswa\typeout
        {IJCGA warning: optional citation argument
        ignored: `#2'} \fi}}

\def\fnt#1#2{\footnotetext{\kern-.3em
        {$^{\mbox{\sevenrm #1}}$}{#2}}}

 1
 1
 1

\font\tenbf=cmbx10
\font\tenrm=cmr10
\font\tenit=cmti10

\font\ninerm=cmr9

\begin{document}
\hfill IFT 94/23

\vspace{5mm}
\centerline{\tenbf TESTING THE $WW\gamma$ COUPLING AT $e^+e^-$
COLLIDERS\footnote{Presented at the Joint US-Polish Workshop ''Physics from
Planck Scale to Electroweak Scale'', Warsaw, September 21-24, 1994}}
\vspace{0.8cm}
\centerline{\tenrm Jan Kalinowski\footnote{Supported in part by the Polish
Committee for Scientific Research Grant 2 P302 095 05}}
\baselineskip=13pt
\centerline{\tenit Institute of Theoretical Physics, Warsaw University}
\baselineskip=12pt
\centerline{\tenit Ho\.za 69, 00681 Warsaw, Poland}
\vspace{0.9cm}
\abstracts{The production of  single photons plus
missing energy
at future $e^{+}e^{-}$ colliders can provide a testing ground for
 non-standard $WW\gamma$
couplings. We show that even with conservative
estimates of systematic errors
there is still considerable sensitivity to anomalous couplings.
}

\vfil
\rm\baselineskip=14pt
In spite of many experimental successes certain aspects of the standard model
still await for direct tests. Among them is the non-abelian sector of the
model and, in fact,  a detailed investigation of gauge boson
self-couplings will constitue
one of the primary physics goals of the next linear colliders.\cite{conf} The
processes with production of gauge bosons in the final state at
future colliders should provide such crucial tests of electroweak
theory.  In particular, they should allow  to improve LEP bounds
on non-Yang-Mills like triple gauge boson vertices.

In this talk we suggest that
the process $e^{+}e^{-}\rightarrow \overline{\nu}\nu\gamma$
 containing a single isolated photon and missing energy
may be used to study the precise structure of
the $WW\gamma$ vertex.\cite{aks} Such final states have been
studied at PETRA\cite{Karel} and at LEP\cite{frits,opal} as
a means of determining the number of light neutrini. At these energies though,
not much sensitivity to anomalous $WW\gamma$ form-factors should be expected,
as s-channel processes mediated by virtual Z exchange also play an
important role. However, at Next Linear Collider (NLC) energies
($\sqrt{s} = 500$ GeV), s channel contributions
become less important and the bulk of the cross-section comes from t channel
W exchange.
The process $e^{+}e^{-}\rightarrow \overline{\nu}\nu\gamma$ from the point of
view of $WW\gamma$ studies has been considered earlier.\cite{blt} The novel
feature of our analysis,
as we will show, is that it is possible to choose cuts which enhance the
sensitivity of the observed cross-sections and differential distributions to
the $WW\gamma$ form-factor.

The advantage of the process $e^{+}e^{-}\rightarrow \overline{\nu}\nu\gamma$
over the most studied\cite{conf} reaction
$e^{+}e^{-} \rightarrow W^{+}W^{-}$
is that the latter, in spite of being a sensitive probe of
non-standard $WW\gamma$ and $WWZ$ couplings, suffers from
the drawback that there
is no obvious way to disentangle the effects of $WW\gamma$ and $WWZ$
form-factors. Hence it is desirable to investigate other channels where
such $\gamma-Z$ interference effects are not present.

The deviations of non-abelian vertices from the standard model couplings,
as was shown by Hagiwara et al.,\cite{HPZH} can be parametrized by
seven possible Lorentz
and $U(1)_{em}$ invariant triple gauge boson form-factors. Requiring
C and P invariance (in the absence of beam polarisation
there is no way to detect CP violating asymmetries if the only particle
detected is a photon of unknown polarisation) then only two,
traditionally denoted by $\kappa$ and
$\lambda$, remain. In the
standard model we have $\kappa =1$ and $\lambda = 0 $. Deviations from the
standard model are then parameterised by $\delta \kappa = \kappa -1 $ and
$\lambda$.  The modified Feynman rules for the $WW\gamma$ vertex may be
obtained from Ref.\cite{yehudai} All other Feynman rules are standard ones.

Due to the complexity of the Feynman rules for the non-standard couplings
it turns out to be convenient to calculate the matrix-element using the
helicity amplitude formalism.\cite{CALKUL,gk} The results are presented in
Ref.\cite{aks}
As a further check we evaluated the helicity amplitudes using
the formalisms of Ref.\cite{Hagizep} and Ref.\cite{gk}
and find excellent numerical agreement
for various values of $\kappa$ and $\lambda$.

Since we assume the electron to be
massless we need to impose a minimum angle cut on the direction of the
outgoing photon to avoid collinear singularities as well as a minimum
energy cut to avoid IR problems. Setting $\theta_{min} = 20^o$ and
$E_{min} = 10$ GeV we find for the standard model
a cross-section of $\sim 1.6$ pb, which at
projected NLC luminosities ($\sim 10$ fb$^{-1}$) represents a sizable
number of events. However, with these cuts alone the sensitivity to
non-standard couplings is rather small because the bulk
of the cross-section comes from initial state soft photon bremstrahlung
which is independent of the non-abelian couplings.
 It is clear that only the more
energetic photons will be sensitive to the anomalous form-factors
which are associated with higher dimensional
operators containing derivative interactions. Therefore
we require that the minumum energy of the photon be 80 GeV. Further
improvement can be obtained by eliminating the background from the Z
exchange graphs, $e^+e^- \rightarrow Z\gamma \rightarrow
\overline{\nu}\nu\gamma$.
This can easily be achieved by means of a simple kinematical
cut because in this case the photon is essentially
monochromatic (in the limit that the $Z$ width is negligible)
with an energy close to half the centre of mass energy.
 Hence we require that the energy
of the photon be less than 180 GeV in order not to reduce
the signal from the $W$
exchange graphs too much.

With these cuts on the photon energy, 80 -- 180 GeV, the cross-section for the
standard model is 0.21 pb, which still leads to an appreciable number of
events at NLC luminosities. Cross-sections for non-standard values of
$\delta\kappa$ and $\lambda$ with the cuts mentioned above are presented in
Fig.1, where we have varied $\delta \kappa$ and $\lambda$
individually
and not simultaneously in order to keep the analysis simple.

As we can see, the process $e^+e^-\rightarrow \overline{\nu}\nu$ is
quite sensitive to the deviations from the standard model. The experimental
limits that can be derived for $\delta\kappa$ and $\lambda$ depend however on
possible experimental and theoretical uncertainties.
Statistical errors are probably quite small given the large number of events
$ {\cal O}( 2000)$. Assuming that there are no experimental
systematic errors, the
main source of theoretical systematic errors lies in unknown higher order
corrections. Note that the higher order corrections discussed in
Ref.\cite{frits}
are those which are dominant on the Z pole, and are therefore not adequate
for our purposes. It is reasonable to assume that the bulk of the
corrections are real and virtual QED corrections which integrated over the
the phase space are probably quite small. However we are restricting
ourselves to a limited region of the total phase space where
radiative corrections may be sizable even though the total
radiative corrections themselves  are small.
Making  the conservative assumption that the
overall systematic uncertainties are  ${\cal O}( 20 \%) $
it is possible to put the folowing discovery bounds
$ -0.6 < \lambda < 0.6$ and $ -0.6 < \delta \kappa < 2.2$
using the cross-section
with the cuts mentioned above as only sensitive variable.

If higher luminosity can be achieved  ($\sim 50$ fb$^{-1}$)
one can be more optimistic\footnote{I thank R. Settles for discussion on
this point} \hspace{1mm} about the systematic uncertainties and
assuming errors to be
${\cal O} (5 \%)$  better limits can be
derived $-0.3 <\lambda <0.3$ and $-0.2<\delta\kappa<0.2$ or $1.2<
\delta\kappa <1.6$.  The discovery limits
derived by Couture and Godfrey,\cite{cg} where similar issues are discussed,
 are even more stringent than ours due to very optmistic assumptions
about the size of theoretical systematic errors.

The total cross section measurement alone will not allow to determine
the parameters $\delta\kappa$ and $\lambda$ unambiguously.
Further refinement
is possible if one considers differential distributions. This is illustrated
in Fig.2a where we have plotted the differential distribution with respect
to photon energy for the standard model and for two values of $\delta
\kappa$.

Although the
cross-sections are almost the same the differential
distributions are different. For the angular distribution, Fig.2b,
 the sensitivity to different values of $\delta\kappa$ is weaker.
Similar effects are observed for $\lambda$ keeping $\delta\kappa=0$.
However, to derive further discovery bounds on the
basis of differential distributions, a detailed  consideration of
detector acceptances and higher order radiative corrections is necessary.

It is interesting to
note that even our conservative results compare favourably
with the bounds obtained by McKellar and He\cite{He}
on the basis of the recent CLEO measurement of
$b \rightarrow s \gamma$.\cite{CLEO}
 Refering to $e^+e^- \rightarrow
W^+W^-$ at LEP II at 190 GeV more stringent
bounds of the order of $\pm 0.5$ for
$\lambda$ and $\delta\kappa$ can be obtained based on theoretically
favored three--parameter fits provided polarization properties
of produced $W$ bosons are fully measured.\cite{bilenky} Errors can be
further reduced either by imposing additional constraints and performing
fits with only one or two free parameters or employing initial beams
polarization.\cite{likhoded} Considerably weaker bounds are deduced
{}from unconstrained multi--parameter fits. However these bounds are set
knowing that one-loop corrections are small leading to smaller
theoretical systematic errors than what we have assumed.
It is also worth pointing out that $W$ pair
production is sensitive to the time-like triple gauge boson form-factors,
whereas the process we are studying  probes the same form-factors in the
space-like region. Therefore both measurements are in some sense complementary.

To conclude, we have demonstrated that
$\sigma(e^{+}e^{-}\rightarrow \overline{\nu}\nu\gamma)$ at projected NLC
energies and luminosities is sensitive to anomalous $WW\gamma$ couplings.
Making conservative estimates for systematic errors involved it is
possible to constrain $\lambda$ and $\kappa$ to lie within the regions
$ -0.6 < \lambda < 0.6$ and $ -0.6 < \delta \kappa < 2.2$.
These bounds can be
improved through knowledge of currently unknown radiative corrections.

\newpage
\nonumsection{Acknowledgements}
\noindent I would like to thank J. Abraham for many discussions and enjoyable
collaboration.

\nonumsection{References}

\newpage
\vspace*{ 7.5cm}
\fcaption{ Cross section for the process $e^+e^-\rightarrow
\nu\bar{\nu}\gamma$ at $\sqrt{s}=500$ GeV (a) as a function
of $\delta\kappa$ for $\lambda=0$
and (b) as a function of $\lambda$ for $\delta\kappa=0$}

\vspace{8cm}
\fcaption{ (a) Energy spectrum and (b) angular distribution
of the photon at $\sqrt{s}=500$ GeV for
the standard model and  for $\delta\kappa=-0.6$ and $2. $ }


\begin{thebibliography}{99}
\bibitem{conf} For a review and references see ``$e^+e^-$ {\it Collisions at
500 GeV: The Physics Potential} '', Proc.\ Munich, Annecy, Hamburg Workshop,
1991-1993, DESY Reports 92-123A+B, 93-123C.
\bibitem{aks} J.K.Abraham, J. Kalinowski, P. \'Sciepko, {\it Phys.\ Lett.}
{\bf B339} (1994) 136.
\bibitem{Karel} K.J.F. Gaemers, R.Gastmans,  F.M. Renard,
{\it Phys.\ Rev.} {\bf D19} (1979) 1605.
\bibitem{frits} F.A. Berends {\em et al.}, {\it Nucl.\ Phys.} {\bf B301}
 (1988) 583.
\bibitem{opal} OPAL Collab., R. Akers{\it et al.}, CERN-PPE/94-105.
\bibitem{blt} G.V. Borisov, V.N. Larin, F.F. Tikhonin, {\it Z. Physik} {\bf
C41} (1988) 287.
\bibitem{HPZH} K. Hagiwara {\em et al.}, {\it Nucl.\ Phys.} {\bf B282}
 (1987) 253.
\bibitem{yehudai} E. Yehudai, SLAC preprint, SLAC-383 (1991).
\bibitem{CALKUL} F.A. Berends, W. Giele, {\it Nucl.\ Phys.} {\bf B294}
 (1987) 700.
\bibitem{gk} J.F. Gunion, Z. Kunszt, {\it Phys.\ Lett.} {\bf B161} (1985) 333.
\bibitem{Hagizep} K, Hagiwara, D. Zeppenfeld, {\it Nucl.\ Phys.} {\bf B274}
 (1986) 1.
\bibitem{cg} G. Couture, S. Godfrey, preprint OICP/C 94-4, UQAM-PHE-94-09.
\bibitem{He}X.G. He, B.H.J. McKellar, preprint UM-P-93/53, OZ-93/14.
\bibitem{CLEO} E. Thorndike (CLEO Collab.), in Proc.\ of the American Physical
Society Meeting, Washington DC, April 1993.
\bibitem{bilenky} M. Bilenky, J.L. Kneur, F.M. Renard, D. Schildknecht,
{\it Nucl.\ Phys.} {\bf B409} (1993) 22.
\bibitem{likhoded} A.A. Likhoded {\em et al.}, preprint IC/93/288,
UTS-DFT-93-22, hep-ph 9309322.

\end{thebibliography}
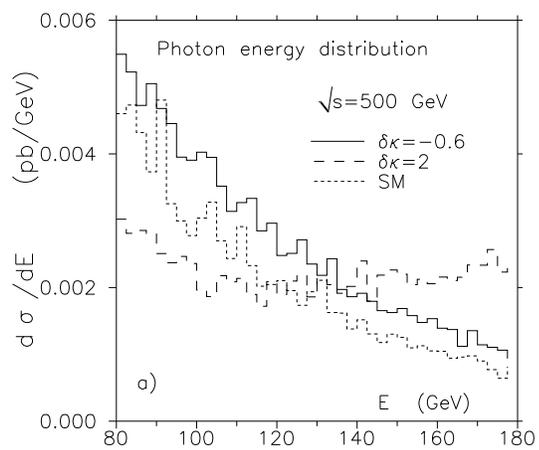
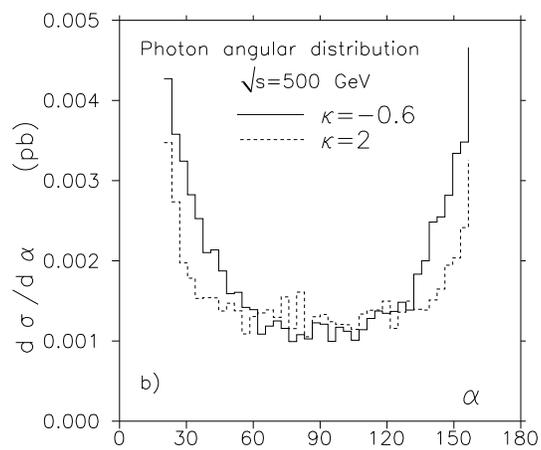
\end{document}